\documentclass[conference]{IEEEtran}
\IEEEoverridecommandlockouts
\usepackage{cite}
\usepackage{amsmath,amssymb,amsfonts}
\usepackage{algorithmic}
\usepackage{graphicx}
\usepackage{textcomp}
\usepackage{xcolor}
\usepackage{csquotes}
\usepackage{listings}
\usepackage{multirow}
\usepackage{tabularx}
\usepackage{enumitem}
\usepackage{comment}
\usepackage{hyperref}

\usepackage{booktabs, multirow} 

\usepackage{caption}
\def\BibTeX{{\rm B\kern-.05em{\sc i\kern-.025em b}\kern-.08em
    T\kern-.1667em\lower.7ex\hbox{E}\kern-.125emX}}
\begin{document}

\title{Requirements Satisfiability with In-Context Learning
}

\author{\IEEEauthorblockN{Sarah Santos, Travis Breaux}
\IEEEauthorblockA{\textit{School of Computer Science} \\
\textit{Carnegie Mellon University}\\
Pittsburgh, Pennsylvannia, United States \\
}
\and
\IEEEauthorblockN{Thomas Norton}
\IEEEauthorblockA{\textit{School of Law} \\
\textit{Fordham University}\\
New York, New York, United States \\
}
\and
\IEEEauthorblockN{Sara Haghighi, Sepideh Ghanavati}
\IEEEauthorblockA{\textit{School of Computing and Information} \\
\textit{University of Maine}\\
Orono, Maine, United States \\
}
}

\maketitle

\begin{abstract}
Language models that can learn a task at inference time, called in-context learning (ICL), show increasing promise in natural language inference tasks. In ICL, a model user constructs a prompt to describe a task with a natural language instruction and zero or more examples, called demonstrations. The prompt is then input to the language model to generate a completion. In this paper, we apply ICL to the design and evaluation of satisfaction arguments, which describe how a requirement is satisfied by a system specification and associated domain knowledge. The approach builds on three prompt design patterns, including augmented generation, prompt tuning, and chain-of-thought prompting, and is evaluated on a privacy problem to check whether a mobile app scenario and associated design description satisfies eight consent requirements from the EU General Data Protection Regulation (GDPR). The overall results show that GPT-4 can be used to verify requirements satisfaction with 96.7\% accuracy and dissatisfaction with 93.2\% accuracy. Inverting the requirement improves verification of dissatisfaction to 97.2\%. Chain-of-thought prompting improves overall GPT-3.5 performance by 9.0\% accuracy. We discuss the trade-offs among templates, models and prompt strategies and provide a detailed analysis of the generated specifications to inform how the approach can be applied in practice.
\end{abstract}


\section{Introduction}

In software engineering, requirements serve to ensure that software is fit for purpose and designed to meet business objectives, legal requirements and stakeholder needs. In their seminal work, Zave and Jackson introduced satisfaction arguments, in which a system's specifications $S$ and domain knowledge or properties $K$ together satisfy the system requirements $R$, written as $S, K \vdash R$. This formulation has been used to motivate goal modeling refinement~\cite{Lam08}, means-end~\cite{MLR+07,LME+12} relationships and traceability~\cite{MWG+16} links, e.g., to show that sub-goals, which refine a parent goal, and domain properties are necessary and sufficient to satisfy the parent goal. 

Satisfaction arguments expressed in logic have the benefit of reducing system behavior, domain properties, and requirements into discrete elements that can be analyzed and evaluated formally. However, these elements can lose the rich description of the argument and associated interpretability that underpins how the requirement is satisfied, and some requirements are difficult to express formally~\cite{Rya93}. Maiden et al.~\cite{MLR+07} and Haley et al.~\cite{HLM+06} extend the formal argument with an informal natural language argument, which Haley et al. call the \textit{inner argument}.

Recent advances in auto-regressive, large language models (LMs) have produced \textit{in-context learning} (ICL), which is the model's ability to recognize a desired task and learn from analogy, often using examples at inference time~\cite{Bro20}. In addition, LMs have been used to perform natural language inference (NLI), including open- and closed-domain reasoning and commonsense reasoning~\cite{RBC+20}. Improvements in NLI have been obtained using chain-of-thought (CoT) prompting, in which the LM is provided one or more reasoning steps prior to generating the answer~\cite{WWS+22}.

In this paper, we study LM applications to generate and evaluate satisfaction arguments. We envision two use cases: (1) a software designer seeks to brainstorm which software features can be used to satisfy a requirement; and (2) given a design description, a designer wishes to check whether that description satisfies a requirement, and further to understand why.

The remainder of this paper is organized as follows: we introduce related work in Section~\ref{section:background}; in Sections~\ref{section:approach} and \ref{section:evaluation}, we present our approach and evaluation method; in Section~\ref{section:results} we present results with discussion in Section~\ref{section:discussion}; in Section~\ref{section:threats} we discuss threats to validity; and we conclude in Section~\ref{section:conclusion}.

\section{Background}
\label{section:background}

We now review related work on natural language inference, in-context learning, satisfaction arguments.

\subsection{Natural Language Inference}

Natural language inference (NLI), which includes Recognizing Textual Entailment (RTE) tasks, is the class of tasks for ``determining whether a natural-language hypothesis can be inferred from a given premise''~\cite{MM08}. Given the text ``A soccer game with multiple males playing,'' a model should confirm that ``some men are playing a sport'' is a valid hypothesis, and that ``the woman kicked the ball'' is a contradiction. Two prominent datasets exist to develop models for NLI, including the Stanford Natural Language Inference (SNLI) corpus~\cite{BAP+15}, and Multi-Genre NLI (MultiNLI) corpus~\cite{WNB18}. The SNLI contains 570K English sentence pairs labeled as entailment, contradiction, and neutral. MultiNLI contains 430K pairs using the same labels, but draws from a wider diversity of domains, including both spoken and written texts. The extended dataset e-SNLI includes explanations to support the predicted label~\cite{CRL+18}. At this time, state-of-the-art neural models exhibit a test accuracy of 93.1\% on SNLI~\cite{WFK+23}, and 90.8\% on MultiNLI~\cite{LOG+19}.

\subsection{In-Context Learning}

Recent advances in auto-regressive, large language models (LMs) have produced \textit{in-context learning} (ICL), which is the model's ability to recognize a desired task and learn from analogy, often using examples at inference time~\cite{Bro20}. For many benchmark natural language processing tasks, supervised deep learning has been the state-of-the-art, often requiring thousands and tens of thousands of training examples~\cite{LYF+23}. With ICL, LLMs have been shown to solve some benchmark tasks with no training examples, called \textit{zero-shot learning}, and as few as 1-15 training examples, called \textit{few-shot learning}\cite{ZWF+21,LBM+22}. To achieve these results, a model user writes a natural language \textit{prompt} consisting of an \textit{instruction} that indicates the type of task, a text linearization called a \textit{template} that includes trigger words and slots that are filled with expressions of human knowledge, and zero or more \textit{demonstrations} or training examples that show the desired input and output~\cite{DLD+23}. 

The choice of which demonstrations to use and their ordering in ICL can have a significant effect on task accuracy by as much as 30\%~\cite{ZWF+21, LBM+22}. In addition, LMs have been shown to exhibit a number of biases in classification tasks when using demonstrations: \textit{majority label bias}, in which LMs choose the most common label among demonstrations; \textit{recency bias}, in which LMs choose the most recent label from the last demonstration; and \textit{common token bias}, in which LMs prefer to output tokens that are more common in their pre-training data~\cite{ZWF+21}. These biases can be mitigated to some extent by ``soft'' prompt tuning, in which the prompt author chooses demonstrations with a near-equal balance of label classes and by ordering the demonstrations to evenly distribute labels.

An instance of ICL in which the demonstration exemplars include the question, the intermediary steps to complete before obtaining the answer, and the answer is called chain-of-thought (CoT) prompting~\cite{WWS+22}. Few-shot CoT prompting has shown improved performance in benchmark NLP tasks for arithmetic, symbolic and commonsense reasoning~\cite{KGR+22}. Zero-shot CoT prompting is used in the absence of exemplars using two-stage prompting: first, the model is prompted to generate the intermediary steps to reach an answer; next, the steps are amended to the first prompt to generate the answer. In arithmetic reasoning, zero-shot CoT performs better than zero-shot prompting, but performs ~10\% worse than few-shot prompting when using Google's PaLM 540B model. There are numerous variations on CoT, including plan-and-solve~\cite{WXL+23}, self-consistency~\cite{WWS+23}, interleaved CoT~\cite{TBK+23} and tree-of-thought prompting~\cite{YYZ+23}.

Instruction tuning is a fine-tuning procedure that improves LM performance when responding to human instructions in natural language~\cite{OWJ+22}. The procedure uses training samples that fit into instructional categories, including brainstorming or making lists, generating narrative and text summarization, to name a few. Several commercial base models have been instruction tuned or provide instruction-tuned variants, including Gemini, Claude, GPT-3.5 and GPT-4. In addition to fine-tuning, prompt authors can divide the prompt template into regions using trigger words to improve performance~\cite{WBZ+22}. Chain-of-thought prompting has been shown to improve performance in multi-hop reasoning tasks, wherein multiple facts must be reasoned about at once or when facts are inferred from other facts in series~\cite{WWS+22}.

Large language models (LMs) are not impervious to failure. LMs have exhibited misdirection, which is the failure to follow instructions~\cite{OWJ+22}. LMs have reported falsehoods, misinformation, and so-called hallucinations~\cite{LHE22}, and they have exhibited gender and racial bias toward others~\cite{NVB+20} and toxicity~\cite{GGS+20}. Moreover, they can exhibit \textit{sycophancy} when they respond with the bias of their users, even if those biases are extreme or incorrect~\cite{WHL+23}. \textit{Alignment} is a broader effort to improve LM performance and reduce these unwanted effects~\cite{ABC+21}, and several commercial models, including Gemini, Claude, GPT-3.5, and GPT-4, have been fine-tuned to this end.

\subsection{Satisfaction Arguments}

Zave and Jackson first introduce \textit{satisfaction arguments} using the formula $S, K \vdash R$, in which system specifications $S$ and domain knowledge and properties $K$ taken together must be sufficient to satisfy requirements $R$~\cite{ZJ97}. In goal modeling, a satisfaction argument is expressed as $SubGoals, DomProps \vdash ParentGoal$, in which satisfying the sub-goals and domain properties is sufficient to satisfy the parent goal~\cite{Lam08}. In goal modeling notation, sub-goals are linked to a parent goal through a refinement relationship, and thus satisfaction arguments can be chained together to trace the satisfaction of an organization's business objectives down to a system's low-level requirements~\cite{Lam08}. 

Maiden et al.~\cite{MLR+07} extended the \emph{i*} diagram notation by attaching satisfaction arguments to the means-end link in the notation. The argument is constructed from S, K and R (above) in addition to an informal, natural language argument justifying the satisfaction of $R$. Lockerbie et al. extended this work to reuse satisfaction arguments across socio-technical system boundaries~\cite{LME+12}.

In addition to refinement and means-end links, Murugesan et al. define traceability as a relationship between target and source artifacts~\cite{MWG+16}. The target artifacts $\Sigma$ implement a system's behavior, such as code, and the source artifacts  $\Delta$ are requirements. Thus, they write $S \vdash_{r} r$ for $S \subset Sigma$ and $r \in \Delta$ when $S$ satisfies the requirement $r$.

Finally, Haley et al. structure satisfaction arguments in two parts: a formal \textit{outer argument} expressed in logic that describes what system behavior entails the satisfaction of the requirement, e.g., $B \vdash R$, which means that requirement $R$ is satisfied by the system behavior by the logical formula $B$; and an informal \textit{inner argument} that supports the claims in the outer argument~\cite{HLM+06}. They propose using causal logic to express the outer argument and Toulmin-style arguments~\cite{Tou58} to structure the outer argument.

\section{Approach}
\label{section:approach}

We first introduce the application domain, before describing our approach to requirements satisfiability using in-context learning, which proceeds in three phases: (1) manually extract and summarize knowledge from authoritative sources for use in natural language inference; (2) generate specifications from public app descriptions; and (3) evaluate requirements satisfaction arguments using the summarized knowledge and generated specifications. The complete replication package is available online.\footnote{\url{https://github.com/cmu-relab/req_sat}}

\subsection{Application Domain}

We made the following assumptions in choosing the application domain. (1) The domain knowledge should be authoritative, reusable, described in general terms independent of any one specification and, when combined with  a specification, the knowledge should be sufficient to infer satisfaction. (2) While specifications can be represented formally (e.g., using logic or graph theory), we assume only informal, natural language descriptions of software. (3) The requirements are mathematically verifiable by satisfying a logical formula consisting of propositions. When true, each proposition corresponds to the satisfaction of an individual requirement described in natural language. (4) Arguments must justify why a requirement is or is not satisfied by drawing connections between the specification and the knowledge about the requirement in question.

Herein, specifications are presented as natural language scenarios with a list of design practices that describe how an application processes personal data. We ask whether the scenario satisfies eight requirements of consent under multiple articles of the EU General Data Protection Regulation (GDPR), including Articles 4(11), 6(1), and 7(4). The knowledge used to check satisfaction is derived from requirement-specific guidance~\cite{EDP20} published by the European Data Protection Board (EDPB), which is the formal body in charge of ensuring the consistent application and enforcement of data processing law in the European Economic Area.

In addition to meeting our assumptions, this problem has additional constraints. While authoritative, the knowledge is limited and dated: the EDPB provides 30 pages of guideline description, including 24 examples, that were authored in May 2020~\cite{EDP20}. From this information, organizations must draw inferences about consent requirement satisfaction. Systems for computing satisfaction must be updated as new knowledge is discovered or created, either by new or amended guidelines, regulatory enforcement actions, or by legal cases. Presently, there are few legal cases or enforcement actions on this topic. We identified only one landmark legal case in Case C-252/21, where the Court of Justice of the European Union (CJEU) found that Meta Platforms violated the consent requirement of freely given by bundling unnecessary advertising practices with other platform data practices~\cite{CJ23}. Similarly, there are few enforcement actions. In our review of 235 enforcement actions~\footnote{https://www.enforcementtracker.com/} decided between 30 June 2023 and 30 December 2023, we found only 50 cases that cover violations of consent articles, among which only 26\% of these cases are likely judgments resulting from software design issues. Due to the limited authoritative ground truth, LMs offer an advantage in utilizing retrieval augmented generation~\cite{LPP+20} to incorporate authoritative regulatory guidance and chain-of-thought prompting~\cite{WWS+22} to make natural language inference explicit.

\subsection{Knowledge}
\label{subsection:knowledge}

Knowledge about the domain and environment are needed to decide if a specification satisfies a requirement. We manually extracted the knowledge and requirements from the guidance document entitled ``Guidelines 05/2020 on consent under Regulation 2016/679, Version 1.1'' that was adopted on 4 May 2020. The document consists of 30 pages and 24 examples that describe scenarios in which systems satisfy or do not satisfy a given property.

We selected all eight requirements from Section 3 plus a ninth requirement from Section 5 of the guidance document to provide breadth in the kinds of phenomena covered. This includes requirements covering how consent is requested, the scope of data practices covered by consent, the scope of information provided to the data subject, and their access to withdrawal. We summarized the guidance in the following abbreviated rubric:
\begin{itemize}[leftmargin=0.4cm]
    \item \textbf{Freely Given (F):} Consent is freely-given, if it exhibits all of the following: (1) is not presented to a data subject by a data controller with a power imbalance; (2) is not conditioned on accepting other terms, and not bundled; (3) is granular; and (4) yields no detriment.
    \begin{itemize}
        \item \textbf{Power Imbalance (P):} Power imbalance generally occurs when the data controller is a public authority or employer, although other cases may arise. For a consent to be freely given in the presence of a power imbalance, the controller must demonstrate that there is no detriment when consent is refused or later withdrawn.
        \item \textbf{Conditionality (C):} If the purpose for processing a data type is bundled with other contract terms, or if the data subject is otherwise compelled to consent, then it is conditional and is not freely given. Conditionality only applies if the requested data is unnecessary to perform the contract. Contracts include end-user agreements, terms of use, and terms and conditions.
        \item \textbf{Granular (G):} Data subjects should be free to choose which purpose they accept, rather than having to consent to a bundle of processing purposes.
        \item \textbf{Detriment (D):} The controller needs to demonstrate that it is possible to refuse or withdraw consent without detriment, including no deception, intimidation, coercion, or significant negative consequences. Gray Area: permissible incentives, which means a controller can use an incentive that is only obtainable if the data subject consents. This incentive is not viewed as a detriment to refusing to consent. Refusal to consent or withdrawal should not lead to a diminished product or service.
    \end{itemize}
    \item \textbf{Specific (S):} The processing of data is limited to specific purposes and will not be processed for other purposes, the consent is granular, and the information presented to obtain consent describes the consent and not other unrelated matters. 
    \item \textbf{Informed (I):} A design description must indicate that a data subject is informed prior to the collection of their data, and at minimum identify (a) the data controller’s identity, (b) the purpose of each processing operation, (c) what type(s) of data will be collected and used, (d) the existence of the right to withdraw consent, (e) information about the use of the data for automated processing, and (f) about the risks due to transfers to countries without adequacy decisions or safeguards.
    \item \textbf{Unambiguous (U):} Consent must be provided through a clear, affirmative action, which may be a written, oral or electronic means.
    \item \textbf{Withdrawal (W):} The data subject can withdraw consent as easily as they gave it, and at any given time.
\end{itemize}

For each requirement definition in the rubric, we wrote a corresponding requirement statement in the optative mood. Requirements are generally written in the \textit{optative mood}, which describes what we desire to be true, whereas the \textit{indicative mood} describes what we assert to be true~\cite{Jac95}. To check if a specification satisfies a requirement, we chose to write the requirements in the indicative mood, because this mood reflects what is or is not true of the specification, as opposed to what could be true, e.g., after modifying the specification or considering a hypothetical. The eight requirements are presented below:

\begin{enumerate}
    \item[P:] There is a power imbalance between the data subject and the data controller.
    \item[C:] The purpose for data processing is bundled with other contract terms, such as user agreements, terms of use, or terms and conditions.
    \item[G:] The data subject can choose which data processing purposes they accept.
    \item[D:] The data subject can refuse or withdraw consent and incur no detriment.
    \item[S:] Data processing is limited to specific purposes.
    \item[I:] The data subject is properly informed.
    \item[U:] Consent is be provided through a clear, affirmative action by the data subject.
    \item[W:] The data subject can withdraw consent as easily as they gave it.
\end{enumerate}

The guidelines indicate that the freely given requirement $F$ is a composition of four refinements, which we formalize in a sub-formula: $\neg P \wedge \neg C \wedge G \wedge \neg D$. Substituting this sub-formula for $F$, we observe that consent validity simplifies to: $\neg P \wedge \neg C \wedge G \wedge \neg D \wedge S \wedge I \wedge U \wedge E \wedge B \wedge W$. In the remainder of this paper, we refer to the eight non-refined requirements, which excludes freely given. Decomposing complex questions into more narrowly focused sub-questions is necessary to reduce the likelihood that the LM will miss relevant details during generation in the satsifiability task, and has been a performant prompting tactic in prior work~\cite{DGS+22, KTF+23, ZSH+23}.

We validated the above rubric by asking three investigators to decide whether the EDPB guidance examples satisfy each of the eight requirements. Among the 24 examples, six examples were excluded from this study because they were either redundant or did not describe a scenario with a target system (e.g., they restated content from recitals). Next, we sanitized the examples to remove any conclusory or justifying language indicating whether a requirement was satisfied and the rationale for satisfaction. For each of the requirements, coders were asked to assign one of the following codes: YES, if the scenario satisfies the property; NO, if the scenario does not satisfy the property; MIS, if the scenario is missing information needed to evaluate the property's satisfaction; or IDK, if the example contains relevant information, but the rubric is unclear. After each coder independently coded all 18 examples, Cohen's Kappa was 47\%. Next, the coders met to discuss their disagreements, after which Kappa rose to 100\%. 

\subsection{Specifications}
\label{subsection:scenarios}

As described in Section~\ref{subsection:knowledge}, the number of authoritative specifications with ground truth labels is limited to 18 examples. To improve external validity, we use the LM to generate specifications from publicly available mobile application (app) descriptions. This has the advantage that the specifications are grounded in real-world applications and we can experimentally control the breadth and diversity of application behaviors. This step also simulates how developers can use LMs to ``brainstorm'' and discover design practices that do and do not satisfy requirements.

The generation process consists of a pipeline of LM tasks, in which the output of an upstream task becomes the input to a downstream task. To describe this process formally, let $I$ be an instruction to an LM that indicates the task type, and let $T$ be a template function that maps one or more text values $(v_{1}, v_{2}, ... v_{n})$ to slots in a text linearization, yielding a slot-filled text $t$. A prompt is the concatenation of one or more strings, expressed using semi-colon, e.g., the prompt $P = I; T(v_{1})$ for an instruction $I$ and a text $t$ based on the template $T$ that had one slot filled by the value $v_{1}$. 

For each task, we used LangChain v0.0.344 and the OpenAI API with the \textit{gpt-3.5-turbo-1106} model and parameters $temperature = 0.7$, $top\_p = 1.0$. We chose this model for the lower pricing and larger context window of 16,385 tokens. The temperature and $top\_p$ parameters control token sampling during generation. Lower temperature gives preference to higher probability tokens, thus reducing randomness and increasing focus across generations, whereas higher temperature generally increases randomness. The $top\_p$ is an alternative to temperature that use nucleus sampling by choosing a subset of tokens with a cumulative probability mass above the $top\_p$ threshold. In our study, we effectively ignore nucleus sampling and chose a moderately lower temperature $0.0 < 0.7 < 2.0$.

We begin with the top 50 mobile application (app) descriptions reported separately by the Google Play and Apple App stores for each of the 27 EU member states. This initial set yields 2,482 unique Apple App apps and 1,559 unique Google Play apps. Next, we randomly selected 200 app descriptions from each app store dataset. The generation process continues through the following steps:

\begin{enumerate}[leftmargin=0.4cm]
    \item For each app description $a_{i} \in A$ for $0 < i \leq |A|$, summarize $a_{i}$ into a one sentence summary $z_{i}$, using the prompt $P_{1} = I_{1}; T_{1}(a_{i})$
    \item For each app description $a_{i} \in A$, extract a list of data practices $D_{i}$, using the prompt $P_{2} = I_{2}; T_{2}(a_{i})$
    \item For each data practice $d_{i,j} \in D_{i}$, identify a list of candidate data types $t_{i,j}$ likely used by the practice, using the prompt $P_{3} = I_{3}; T_{3}(d_{i,j})$
    \item For each summary $z_{i}$, data practice $d_{i,j}$ and list of data types $t_{i,j}$ for this practice, write a brief scenario $c_{i}$ using the prompt $P_{4} = I_{4}; T_{4}(z_{i}, d_{i,j}, t_{i,j})$
    \item For each scenario $c_{i}$, requirement $r_{m} \in R$ and domain knowledge $k_{m}$ defining the requirement, choose one of two augmentations for $c_{i}$ at random: a) assume $r_m$ is true, then generate a list of satisfactory design practices using the prompt $P_{5}^{+} = I_{5}^{+}; T_{5}(c_{i}, k_{m}, r_{m})$; or b) assume $r_m$ is false, then generate a list of dissatisfactory design practices using the prompt $P_{5}^{-} = I_{5}^{-}; T_{5}(c_{i}, k_{m}, r_{m})$. For each list of design practices $X_{i}$ generated using scenario $c_{i}$, let $S_{m}^{+}$ be the set of pairs $(c, X)$, called a specification, that were generated using prompt $P_{5}^{+}$ to satisfy requirement $r_{m}$, and let $S_{m}^{-}$ be the set of specifications generated using prompt $P_{5}^{-}$ to not satisfy requirement $r_{m}$.
\end{enumerate}

At step 3 above, the 200 app descriptions for Apple App and Google Play yielded 1,637 and 2,148 triples consisting of a shared summary, a single data practice, and data types associated with the practice, respectively. As inputs to generating app scenarios in step 4, we randomly sampled 200 triples for each app store. Step 4 yielded 400 scenarios in total. Below is an example scenario generated from steps 1-4:

\begin{displayquote}
\textbf{Summary}: The mobile app is a platform for watching and creating short-form videos that are personalized to your interests, with easy-to-use tools for editing and adding effects, music, and filters.

\textbf{Data Practice}: The user watches short-form videos on TikTok.

\textbf{Data Types}: ['watch history', 'liked videos', 'comments', 'user profile information']

\textbf{Scenario}: The app provides a user with a personalized video feed that offers a wide range of entertaining and inspiring short videos. It also offers simple-to-use tools for creating original videos with special effects, filters, and music. The user is able to edit their own videos using the app's features. The app utilizes videos to enable this functionality.
\end{displayquote}

In step 5, we extend each scenario with a list of design practices generated by the model that collectively satisfy or dissatisfy one of the eight requirements described by the rubric in Section~\ref{subsection:knowledge}. For an even distribution across app stores, requirements and satisfaction states, this yields 384 total specifications down-sampled from the 400 scenarios randomly sampled in step 4. Next, we generate design practices $D$ using deduction expressed by the formula $K, S, R \vdash D$. Because LMs can be fine-tuned to follow instructions~\cite{OWJ+22} and aligned to limit undesired behaviors~\cite{ABC+21} (e.g., toxicity~\cite{GGS+20}, social bias~\cite{NVB+20} and hallucinations~\cite{LHE22}), and because $K$ can overlap with the pre-training data~\cite{Bro20}, which can cause related but irrelevant information to ``leak'' into the model output, we conducted a series of design iterations to shape the answers, called answer shaping~\cite{LYF+23}. Answer shaping includes changes to the input to exclude undesirable output and retain desirable output.


Below is an example list of design practices generated by step 5 for the consent requirement of specificity. Each specification consists of the design practice list appended to the end of the generated scenario.

\begin{enumerate}[leftmargin=0.4cm]
\item The app clearly and specifically states the purposes for which the data will be processed, such as providing a personalized video feed and offering video creation tools with special effects, filters, and music. This ensures that the data processing is limited to these specific purposes.
\item When the user first installs the app, they are presented with a consent request that is granular and allows them to choose whether they consent to each specific purpose separately. For example, they can choose to consent to the personalized video feed but not to the video creation tools, or vice versa. This granular consent ensures that the user has a choice in relation to each specific purpose.
\item The information presented to obtain consent focuses solely on the specific purposes for which the data will be processed. The consent request does not include unrelated matters, such as promotional offers or unrelated data processing activities. This clear separation of information ensures that the consent request describes the consent and not other unrelated matters.
\item The app does not process the user's data for purposes other than those specifically stated. It does not use the data for targeted advertising or share it with third parties without the user's explicit consent. This adherence to purpose specification safeguards against function creep and ensures that the data processing remains limited to the specific purposes for which the user has given consent.
\end{enumerate}

We evaluated the generated dataset in two steps. First, two investigators involved in designing the generation process manually reviewed the 400 sampled scenarios to validate that the data types are relevant to the data practices and that the practices are relevant to the summary. This step identified 19 scenarios for removal due to practices outside app scope, no practices, and misinterpretation of a fictional game. Next, one of these investigators reviewed the generated design practices for logical consistency with whether the practices were expected to satisfy the requirement. We report and discuss the results of this review in Section~\ref{section:discussion}. 


Finally, two investigators not involved in the generation process design used the rubric described in Section~\ref{subsection:knowledge} to classify each generated specification by whether it satisfies or does not satisfy the given requirement and to record their justification for their classification. They first coded 16 specifications before meeting to discuss the process. At this stage, the initial inter-rater reliability calculated using Cohen's Kappa~\cite{Coh60} was $K$ = 50\%, which is considered a moderate agreement~\cite{LK77}. Next, they coded the remaining specifications independently, yielding a Kappa $K$ = 76\%, which is considered substantial. The two investigators further discussed and re-coded five remaining disagreements to yield $K$ = 91\%, which is considered \emph{almost perfect}. 

Among the disagreements, most occur in specifications related to the ``conditionality'' and ``power imbalance'' requirements. In specification SCR-A003, the design practice list states the "Conditionality" to be false, as the data processing now requires the data subject (user) to accept terms and conditions.'' The first investigator decided that conditionality is true because the user must agree to the terms and conditions before using the app, while the second investigator agreed with the quoted conclusory statement and marked conditionality as false. To mitigate these disagreements, the investigators agreed to ignore the conclusory statements and make their own judgement.

The final dataset consisting of the reconciled disagreements were used as the ground truth to evaluate the requirements satisfiability phase, which we now discuss.

\subsection{Satisfiability}
\label{subsection:satifaction}

We check whether a specification $s$ satisfies a requirement $r_{m}$ using knowledge $k_{m}$ specific to the requirement and the prompt $P_{R} = I_{R}; T_{R}(s, k_{m}, r_{m})$. When checking satisfaction, we retrieve the knowledge and requirement from a database, which can be updated over time. The knowledge $k$ is comprised on a requirement name and definition of that requirement, and the specification $s = (c, X)$ is comprised of the app scenario $c$ and list of design practices $X$ for the given app. To satisfy a requirement $r$, it must be true that no design practice $x \in X$ dissatisfies $r$; and to dissatisfy $r$, it must be true that at least one $x \in X$ dissatisfies $r$. To test whether LMs are robust reasoners when requirements statements are inverted, we check satisfiability separately using the requirement $r_{m}$ and the inverted requirement, written $r_{m}^{-}$. For example, the requirement for specificity is written as ``data processing is limited to specific purposes'' and the inversion is written as ``data processing is not limited to specific purposes.'' Therefore, we can check for consistency such that $s, k \vdash r$ is not true, if and only if $s, k \vdash \neg r$ is true.

\subsubsection{Instructions and Templates} Prompt performance has been shown to vary widely based on the choice of words in the instruction and template~\cite{SRR+22, ZWF+21}. This variance is attributed to the word distributions learned during the model pre-training and to how in-context learning uses token sequences to select the task at inference time. Thus, we examine two different templates by varying the prompt vocabulary. In Listing~\ref{listing:tempR}, we present a \textit{requirements instruction and template} $I_{R}; T_{R}$. In this template, the answer is missing and the model is instructed to complete the answer with True or False. In Listing~\ref{listing:tempG}, we present a \textit{generic instruction and template} $I_{G}; T_{G}$ replaces the ``specification'' and ``requirement'' trigger words from $I_{R}; T_{R}$ with ``scenario'' and ``statement,'' respectively. Apart from these changes in the trigger words and in how to choose a response, the two templates are identical.

\begin{minipage}{.95\linewidth}
\begin{lstlisting}[language=HTML,caption={Satisfiability Template R},label={listing:tempR},basicstyle=\scriptsize\ttfamily]
Definition of {req_name}: {definition}
    
Read the following specification. If the specification satisifies the requirement based on the definition, above, respond with True, otherwise respond with False. Do not comment or elaborate.

Specification: {scenario} {design_practices}

Requirement: {requirement}

Answer: 
\end{lstlisting}
\end{minipage}

\begin{minipage}{.95\linewidth}
\begin{lstlisting}[language=HTML,caption={Satisfiability Template G},label={listing:tempG},basicstyle=\scriptsize\ttfamily]
Definition of {req_name}: {definition}
    
Read the following scenario, and decide if the given statement is true or false based on the definition, above. Respond with True or False. Do not comment or elaborate.

Scenario: {scenario} {design_practices}

Statement: {requirement}

Answer: 
\end{lstlisting}
\end{minipage}

\subsubsection{Chain-of-Thought Prompting} Prompts that involve reasoning over multiple hops or facts have shown improved performance using Chain-of-Thought (CoT) prompting in natural language inference tasks~\cite{WWS+22}. In CoT prompting, a prompt is augmented with one or more examples that demonstrate step-by-step reasoning. The model then completes a reasoning task prior to providing an answer to a given question. The key idea is that the generated reasoning reduces the probability of misdirection when completing the answer. To further narrow the focus in this task, we chose to apply CoT directly to the list of design practices and to exclude the scenario. In Listing~\ref{listing:chain_thought}, we introduce prompt $P_{CoT} = I_{CoT}; T_{CoT}(X, k_{m}, r_{m}, E)$ that checks whether a list of design practices $X$ satisfy requirement $r_{m}$ given knowledge $k_{m}$. Based on an early evaluation of $P_{G}$, we based this template on the generic template $T_{G}$. The template $T_{CoT}$ embeds a list of training examples $E$ in which each example $e \in E$ consists of a list of design practices, a requirement, an answer to whether the design practices satisfy the requirement, and a rationale justifying the answer (see sub-template $T_{CoT}^{E}$ in Listing~\ref{listing:chain_example}). To separate the examples, we use the trigger word \# END, and to separate the question and answer components we use the trigger word \#\#\#. These examples are drawn from the ground truth dataset and kept separate from the data used for evaluation.

\begin{minipage}{.95\linewidth}
\begin{lstlisting}[language=HTML,caption={Chain-of-Thought Template},label={listing:chain_thought},basicstyle=\scriptsize\ttfamily]
Definition of {req_name}: {definition}

Read the following example scenarios and observe the rationale and answer about whether the statement is true or false. For the last scenario and statement, decide if the statement is true or false based on the definition, above. Respond by completing the Rationale and Answer using the same format. Do not elaborate.

{examples*}

Scenario: {design_practices}

Statement: {requirement}

###

Rationale:  
\end{lstlisting}
\end{minipage}

In Listing~\ref{listing:chain_thought}, the \texttt{\{examples*\}} slot is filled with one or more examples generated by the sub-template show in Listing~\ref{listing:chain_example}. The slot fillers in each sub-template are drawn from the ground truth data hold-out dedicated to training.

\begin{minipage}{.95\linewidth}
\begin{lstlisting}[language=HTML,caption={Chain-of-Thought Sub-Template},label={listing:chain_example},basicstyle=\scriptsize\ttfamily]
Scenario: {design_practices}

Statement: {requirement}

###

Rationale: {rationale}

Answer: {answer}

# END
\end{lstlisting}
\end{minipage}

As described in Section~\ref{section:background}, the choice of which demonstrations to use and their ordering can impact accuracy by as much as 30\%~\cite{ZWF+21, LBM+22}. This is due in part to \textit{majority label bias}, in which LMs choose the most common label among demonstrations, and \textit{recency bias}, in which LMs choose the most recent label from the last demonstration~\cite{ZWF+21}. These effects are reduced as model size increases. Regardless, to mitigate any possible effects of these biases, we divide the training examples into subset $E_{m}^{+}$ for those examples that demonstrate when requirement $r_{m}$ is satisfied, and into subset $E_{m}^{-}$ for those examples that demonstrate when requirement $r_{m}$ is not satisfied. Next, we check the satisfiability of $r_{m}$ for a previously unseen list of design practices $X$ by presenting demonstrations in the ordering $e_{0}^{+}; e_{0}^{-}; ... ; e_{n}^{+}; e_{n}^{-}$ for $e_{i}^{+} \in E_{m}^{+}$ and $e_{i}^{-} \in E_{m}^{-}$ and $0 < i \leq min(|E_{m}^{+}|, |E_{m}^{-}|)$. This ordering ensures an equal number of examples from each class (satisfies and does not satisfy) and distributes the classes evenly in the order. Finally, if we are testing an inverted requirement, then we invert the requirement and answer in all of the selected examples.

\section{Evaluation}
\label{section:evaluation}

We evaluate the efficacy of checking requirements satisfiability by the proportion of correct LM responses ($c$) to the total LM responses ($t$), which is equal to $c / t$, called \textit{accuracy}. To check if a response is correct, we first check if the case-insensitive response is in the set \texttt{(True, False)}, which we call a \textit{uniform response}. This check is necessary because generative LMs can produce answers that are not limited to a discrete set, despite considerable answer shaping. If the response is not \texttt{True} or \texttt{False}, which we call a \textit{non-uniform response}, we next try to match the response to the regular expression \texttt{/Answer: (True|False)/} based on the template design and trigger word \texttt{Answer}, and to the expression \texttt{/The (requirement|statement) ("?.+?"?\textbackslash s)?is (true|false)/}, which is a common elaboration featured by the model that in some cases includes the requirement restated in quotes. This approach is helpful to detect correct responses when the LM generates additional commentary and embeds the answer in this commentary. If the regular expressions do not match a response, which we call a \textit{non-parsable response}, we count this response as incorrect. If the expression matches one or more times, then we accept the last match as the predicted answer and check the predicted answer against the expected answer in the ground truth dataset. We report the number of non-uniform responses that do not match the above patterns for each experiment in Section~\ref{section:results}.

In addition to computing overall accuracy, we are interested in how accuracy varies by requirement type, how the effects of the polarity of the scenario affect accuracy, which is whether the scenario satisfies or dissatisfies the requirement, and how the polarity of the requirement affects accuracy, which is whether the requirement is written in terms of what the system does or does not do. Finally, we examine whether running a prompt 10 times and taking the majority response (vote) affects accuracy, which is called self-consistency~\cite{WWS+23} and has shown promise in prior work~\cite{LAC+21}. To this end, we conduct experiments to answer the following research questions:

\begin{itemize}[leftmargin=1.0cm]
    \item[\textbf{RQ1}:] How does accuracy vary by requirement type?
    \item[\textbf{RQ2}:] How does the scenario polarity affect accuracy?
    \item[\textbf{RQ3}:] How does the requirement polarity affect accuracy?
    \item[\textbf{RQ4}:] How does majority response affect accuracy?
\end{itemize}

To evaluate overall accuracy and answer the research questions, we used the ground truth dataset described in Section~\ref{subsection:scenarios}. We sampled the dataset to ensure half of the specifications were generated from the 1,637 Apple App scenarios, and the other half from the 2,148 Google Play scenarios. In addition, we sampled to ensure an even distribution across the eight requirements and two satisfaction states, wherein a specification either satisfies or does not satisfy one of the eight requirements. We restricted the sample total to less than 400 specifications, which yields 384 specifications in which each requirement and satisfaction state was replicated 24 times, and each specification represents a different mobile application (i.e., $8 \times 2 \times 24 = 384$). Because Chain-of-Thought prompting requires training examples, we held out 20\% of the ground truth dataset based on a near even distribution across all eight requirements, which yields 75 training samples, and 300 testing samples. We evaluated each satisfiability prompt $P_{6}$ and $P_{7}$ using the same 300 testing samples, which were evenly distributed across requirements.

In-context learning relies on token sampling to generate a model completion. Sampling is a non-deterministic process that can be affected by changing the model $temperature$ or $top\_p$ parameters. This can lead to different responses, even when using the same prompt. Therefore, we checked satisfiability by prompting the model 10 times for each requirement and its inversion to yield a total of $300 \times 2 \times 10 = 6,000$ LM responses per experiment. We report the mean accuracy of the 10 trials for each experiment. For the majority response, we calculate the frequencies for a True and False response, respectively, and choose the response with the highest frequency. If the responses are tied, we randomly choose True or False as the response.

For this task, we used LangChain v0.0.344 and OpenAI API with the \textit{gpt-3.5-turbo-1106} model, which has a 16,385 token context window, and the \textit{gpt-4-0613} model, which has a 8,192 token window. Both models have a pre-training cut-off date of September 2021. In each experiment, we use the same parameters $temperature = 0.7$, $top\_p = 1.0$. 

\section{Results}
\label{section:results}

We now discuss the results based on the approach described in Section~\ref{section:approach}. In Table~\ref{table:overallresults}, we present the mean accuracy for the 10 trials for each experiment and majority response described in the evaluation method in Section~\ref{section:evaluation}. The columns correspond to each experiment, including the requirements $P_{R}$ and generic $P_{G}$ templates, the best performant 1-shot Chain-of-Thought (CoT) template $P_{CoT}$ and both GPT-3.5 and GPT-4 models. The rows present the per-requirement mean accuracy, the accuracy when the specification satisfies the requirement (Spec. true) and dissatisfies the requirement (Spec. false), and when the requirement was written in the normative and inverted tone, the overall accuracy followed by the accuracy when taking the majority response from 10 prompt responses. The highest accuracy for each experiment is presented in \textbf{bold}. All experiments used the same dataset for this evaluation.

The highest overall accuracy for checking requirements satisfaction was 95.6\% using GPT-4 and the requirements template $P_{G}$, followed by GPT-4 with the generic template and closely by GPT-3.5 with 1-shot Chain-of-Thought (CoT) prompting. CoT has the advantage that responses include the rationale or justification for the answer and GPT-3.5 is $10^4$ less expensive to operate. In these experiments, we only observed unparsable, non-uniform responses in the CoT experiment that accounts for 0.002\% of the error in that reported result.

\begin{table}
\centering
\begin{tabular}{l | c c c | c c} 
\toprule
 \multirow{2}{*}{} & \multicolumn{3}{c|}{\textbf{GPT-3.5}} & \multicolumn{2}{c}{\textbf{GPT-4}} \\
& \textbf{$P_{R}$} & \textbf{$P_{G}$} & \textbf{$P_{CoT}$} & \textbf{$P_{R}$} & \textbf{$P_{G}$} \\[0.2cm] 
\midrule
\multicolumn{1}{l|}{Power Imbalance} & 0.876 & 0.849 & 0.914 & \textbf{0.950 }& 0.895 \\ 
\multicolumn{1}{l|}{Conditionality} & 0.761 & 0.741 & 0.722 & \textbf{0.797} & 0.791 \\ 
\multicolumn{1}{l|}{Granularity} & 0.676 & 0.871 & \textbf{1.000} & 0.989 & 0.986 \\ 
\multicolumn{1}{l|}{Detriment} & 0.849 & 0.895 & 0.997 & \textbf{1.000} & \textbf{1.000} \\ 
\multicolumn{1}{l|}{Specificity} & 0.730 & 0.805 & 0.955 & 0.971 & \textbf{0.974} \\ 
\multicolumn{1}{l|}{Informed} & 0.672 & 0.804 & 0.999 & \textbf{1.000} & 0.976 \\
\multicolumn{1}{l|}{Unambiguous} & 0.675 & 0.839 & 0.942 & \textbf{0.962} & 0.934 \\
\multicolumn{1}{l|}{ Withdrawal} & 0.846 & 0.922 & 0.914 & \textbf{0.978} & 0.897 \\  
 \midrule
\multicolumn{1}{l|}{Spec. (true)} & 0.889 & 0.899 & 0.947 & \textbf{0.967} & 0.952 \\ 
\multicolumn{1}{l|}{Spec. (false)} & 0.486 & 0.716 & 0.896 & \textbf{0.932} & 0.888\\ 
 \midrule
\multicolumn{1}{l|}{Req.} & 0.633 & 0.841 & 0.933 & \textbf{0.940} & 0.923 \\
\multicolumn{1}{l|}{Req. (inverted)} & 0.889 & 0.840 & 0.929 & \textbf{0.972} & 0.940 \\
 \midrule
\multicolumn{1}{l|}{Overall Accuracy} & 0.761 & 0.841 & 0.931 & \textbf{0.956} & 0.932 \\
\multicolumn{1}{l|}{Maj. Response} & 0.783 & 0.855 & 0.938 & \textbf{0.959} & 0.931 \\
 \bottomrule
\end{tabular}
\vspace{0.2cm}
\caption{Mean Accuracy over 10 Experimental Trials}
\label{table:overallresults}
\vspace{-0.4cm}
\end{table}

RQ1 asks ``How does accuracy vary by the requirement type?'' To answer RQ1, we examine the accuracy for each requirement. In Table~\ref{table:overallresults}, Conditionality was overall the weakest performing satisfiability check across all models and prompt types. Conditionality was also a requirement that the human evaluators identified the most disagreements over. In weaker performing experiments, Informed is a close second in weakest performance, whereas GPT-4 performs very well for this requirement. Notably, CoT reaches near optimal performance for Informed reaching 99.9\% accuracy.

The question RQ2 asks ``How does the scenario polarity affect accuracy?'' Within the ground truth dataset, 192 specifications were generated to satisfy a requirement (true), and 192 specifications were generated to dissatisfy a requirement (false). In general, we observe in Table~\ref{table:overallresults} that accuracy falls when the specification does not satisfy the requirement. The highest performance loss is GPT-3.5 with the requirements prompt $P_{R}$, dropping to 48.6\% accuracy as compared to 71.6\% with the generic template. Even the best performing GPT-4 experiments exhibit a 4-7\% difference in performance.

The question RQ3 asks ``How does the requirement polarity affect accuracy?'' In this experiment, the requirement was written as a normative statement describing \textit{what a system should do}, and as an inverted statement describing \textit{what the system should not do}. The requirement template, the normative statement and GPT-3.5 shows the lowest accuracy (63.3\%) with less difference among GPT-3.5 CoT and GPT-4.

Finally, RQ4 asks ``How does the majority response affect accuracy?'' In this experiment, we observe that majority response generally yields an accuracy close to or slightly above (1-3\%) the mean. This observation means this option is preferred, since there are trials among these experiments where the accuracy is below the the mean.

\begin{figure}[htbp]
\centerline{\includegraphics[scale=0.6]{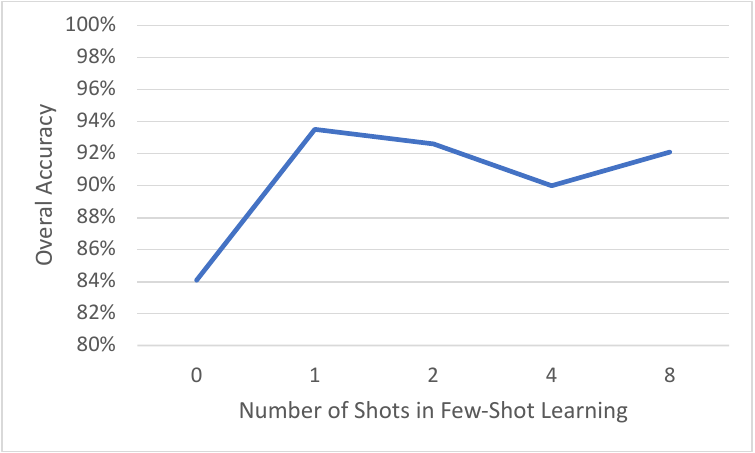}}
\caption{$n$-Shot CoT Accuracy with Generic Template}
\label{figure:nshot_cot}
\end{figure}

In Table~\ref{table:overallresults}, we presented the best $n$-shot CoT experiment, which is $n = 1$. In Figure~\ref{figure:nshot_cot}, we present the overall mean accuracy for the GPT-3.5 generic prompt $P_{G}$ experiment to represent $n = 0$ and the CoT prompt ($P_{CoT}$) experiments for $n = 1, 2, 4, 8$. The one-shot experiment yields the highest overall accuracy and declines up to four-shots before increasing again with eight-shot. We discuss an explanation for this decline in Section~\ref{section:discussion} after performing an error analysis on the unparsable, non-uniform responses.

Table~\ref{table:non-uniform_responses} presents frequencies of non-uniform responses for each experiment. The first column indicates the number of parsable answers, with the remaining columns corresponding to the experiments with each of the requirements ($P_{R}$) and generic prompts ($P_{G}$) tested, the $n$-shot ($n$S) CoT prompt ($P_{C}$) and the two models, GPT-3.5 and GPT-4. The rows correspond to the number of responses out of the total of 6,000 possible responses for each experiment. Nearly all of the CoT template responses were non-uniform and 4-shot CoT showing the highest number of non-parsable responses. The requirement and generic templates used with GPT-4 show zero non-uniform responses. 

\begin{table}
\centering
\begin{tabular}{l c c c c c c c c} 
\toprule
\multirow{2}{*}{\textbf{Ans.}} & \multicolumn{6}{c}{\textbf{GPT-3.5}} & \multicolumn{2}{c}{\textbf{GPT-4}} \\
 & $P_{R}$ & $P_{G}$ & $P_{C}$1S & $P_{C}$2S & $P_{C}$4S & $P_{C}$8S & $P_{R}$ & $P_{G}$ \\[0.2cm] 
\midrule
\multicolumn{1}{l}{\textbf{0}} & 0 & 0 & 7 & 15 & 17 & 48 & 0 & 0\\ 
\multicolumn{1}{l}{\textbf{1}} & 18 & 11 & 6066 & 6034 & 6010 & 6003 & 0 & 0\\ 
\multicolumn{1}{l}{\textbf{2}} & 0 & 0 & 3 & 10 & 11 & 10 & 0 & 0\\ 
\multicolumn{1}{l}{\textbf{3}} & 0 & 0 & 0 & 20 & 1 & 2 & 0 & 0\\ 
\multicolumn{1}{l}{\textbf{4+}} & 4 & 0 & 1 & 0 & 40 & 1 & 0 & 0\\ 
\bottomrule
\end{tabular}
\vspace{0.2cm}
\caption{Non-uniform Responses among 6000 Total}
\label{table:non-uniform_responses}
\vspace{-0.4cm}
\end{table}

\section{Discussion}
\label{section:discussion}

We now discuss and interpret our results.

While 1-shot Chain-of-Thought (CoT) prompting shows improvement over non-CoT prompting with GPT-3.5, we observed a decreasing accuracy with increasing $n$ and a high number of non-uniform responses. The pattern with non-uniform responses shown in Table~\ref{table:non-uniform_responses} for 1-shot CoT is consistent across all $n$ for $n = 1, 2, 4, 8$: most responses are non-uniform, however, the number of unparsable non-uniform responses where zero answers were received increases in proportion to the loss of accuracy. We examined the unparsable responses and observed the model was increasingly ignoring the instructions with phrases such as ``Apologies, but I cannot fulfill that request.'' and in one case ``I'm sorry, but I cannot complete this task as it requires making judgments about legal and ethical matters.'' In 4-shot CoT where overall accuracy was the least, 33/45 unparsable responses exhibit this pattern. 

We attribute these unparsable responses to the practice of \textit{alignment}, which aims to produce models that are helpful, honest and harmless~\cite{ABC+21}. Alignment reduces toxicity in responses and can prevent providing responses that conflict or compete with licensed professional advice, such as legal or medical guidance. Because our chosen application domain is legal compliance, the CoT prompts may have triggered alignment protocols. The models that we studied, \textit{gpt-3.5-turbo-1106} and \textit{gpt-4-0613}, have both been instruction-tuned and aligned according to OpenAI's public disclosures. An alternative to using an aligned model is to use a base model, such as LLaMA2 or \textit{davinci-002}, that has not been instruction-tuned nor aligned. The downside is that instruction-tuned models have shown improved performance over base models on benchmark NLP tasks~\cite{WBZ+22}. That said, aligned models exhibit an ``alignment tax'' that reduces NLP performance~\cite{OWJ+22}.

We conducted an analysis of the total 3,291 generated design practices across 384 specifications to identify errors and other issues in the generation method described in Section~\ref{subsection:scenarios}. Overall, we observed 29 specifications (7.55\%) in which the design practices exhibit the opposite polarity of the intended requirement. Nine of these specifications were generated to fit Conditionality, and the remaining 20 to fit Detriment. We observed 12 specifications with a definition bias (3.13\%), in which certain elements of a definition are emphasized while others are minimized or missing. For example, the practices for Conditionality overemphasize ``accepting terms and conditions'' and under-emphasize ``bundling consent with unnecessary data.'' 

We present summary statistics for eight categories of concern in the generated design practices in Table~\ref{table:practice_review}.

\begin{table}[!htp]\centering
\caption{Frequency of Generation Issues Identified}
\label{table:practice_review}
\begin{tabular}{lrrrrr}
\toprule
Issue Category & Frequency & Percentage \\
\midrule
Property Inversion &29 &7.65\% \\
Legal Nuance &24 &6.3\% \\
Inference and Interpretation &15 &3.9\% \\
Definition Bias &12 &3.1\% \\
Invalid Logical Leaps &6 &1.6\% \\
Inconsistent Design Practices &5 &1.3\% \\
Double Negatives &3 &0.8\% \\
\bottomrule
\end{tabular}
\end{table}

The generated design practices often contained conclusory statements that explicitly refer to whether the requirement is satisfied. We identified 256 specifications (66.67\%) with conclusory statements, among which 66 could be overtly leading. For example, SCR-G028 states ``...the user is consenting to the processing of their personal data for multiple purposes related to enhancing the security of their online accounts. However, since the user cannot choose which specific processing purposes they accept or give separate consent for each purpose, the 'granularity' requirement is not fulfilled.'' Conclusory statements also appear in the EDPB authoritative examples, and were ignored (if they were inaccurate) when creating the ground truth dataset.

We observed that 3.91\% of specifications contained irrelevant practices with a conclusory statement, e.g., in SCR-G124 ``These actions do not directly address the power imbalance between the data subject and the data controller. However, it is important to note that the presence of power imbalance is not negated by these design practices.''Six specifications (1.56\%) exhibited fallacious reasoning, including practice \#5 of SCR-G144 ``The user saves the collage, indicating that the data processing was limited to the specific purpose of creating the collage and not used for other unrelated matters.'' This is unsound, because ``saves the collage'' does not entail ``processing was limited to the specific purpose.'' 

We observed a few instances logical inconsistency. Five specifications (1.30\%) contain logically inconsistent design practices. For example, in SCR-G187, design practice \#5 states ``The user reads and accepts the terms and conditions,'' while practice \#13 states ``The user chooses not to accept the terms and conditions.''Finally, we observed three specifications (0.78\%) with double negatives, including SCR-G080 ``This design practice does not cause 'Detriment' to be false.''

\section{Threats to Validity}
\label{section:threats}

We now discuss threats to validity.

\textit{Construct validity} is the correctness of operational measures used to collect data, build theory and report findings from the data~\cite{Yin03}, and the extent to which an observed measurement fits a theoretical construct~\cite{SCC02}. To reduce this threat, we developed the knowledge and rubric described in Section~\ref{subsection:knowledge} used to generate the design practices and evaluate satisfiability from the authoritative European Data Protection Board consent guidelines~\cite{EDP20}. In addition, we divided the investigators into the process group and evaluation group. The process group evaluated the rubric using the 18 examples described in the guidelines, and the evaluation group evaluated the specifications generated using the rubric. Finally, we seeded every specification using mobile app summaries and data practices extracted from real descriptions of top, most popular apps used in the jurisdiction under which the legal requirements apply.

It is likely that the knowledge used to generate the design practices is incomplete, and that other unforeseen design practices will arise as technology evolves. At present, industry only has 18 authoritative examples from which to evaluate their own designs, and there are few regulatory enforcement actions and few judicial cases to clarify unforeseen situations. Thus, future work should study the effects of extending the knowledge to address new challenges.

In addition, the design practices themselves were generated by the LM in zero-shot setting with only the knowledge to guide the generation. This step yields synthetic data that may not be representative of actual design practices. To mitigate this threat, the process group reviewed the generated design practices for anomalies and inconsistencies. However, they did not evaluate the generated practices using outside knowledge, e.g., using a survey of industrial practices.

\textit{Internal validity} is the extent to which measured variables cause observable effects in the data~\cite{Yin03}. In this study, we created synthetic data consisting of design practices generated by the GPT-3.5 language model using knowledge provided by an authoritative source and the model's pre-training data. In two experiments, we used the same model to check whether the generated specifications satisfy legal requirements described in the same source of knowledge. It's possible that the model's output in the experiments was biased by the model's output in the specification generation step to predict the expected answer. To mitigate this step, we had two investigators not involved in the generation process to independently create expected answers in a ground truth dataset. This dataset was used in the evaluation to ensure independence between generation and satisfiability checking. In addition, we employed a second larger model GPT-4 to evaluate results independently from the model used to generate the specifications.

It is possible that the EDPB guidelines that were used to create the knowledge base, including the 18 examples, were part of the LM pre-training data. The cut-off dates for the GPT-3.5 model \textit{gpt-3.5-turbo-1106} and GPT-4 model \textit{gpt-4-0613} are September 2021. This means performance may worsen or require additional context in cases where the requirements are completely unseen by a different LM.

\textit{External validity} determines the scope of environmental phenomena or domain boundaries to which the theory and findings generalize~\cite{Yin03}. In this study, we examined requirements in a legal domain applied to mobile applications. The requirements cover a range of system behavior described in the specifications, including obligations of the app developer and their firm, how data is processed, how consent is collected and what happens when consent is withdrawn, and how information is presented to app users. Other phenomena not covered by this setting include performance requirements, including concurrency and parallelism in computation, safety critical requirements, and security requirements.

In addition, prior research has shown that prompt performance does not transfer between models in a predictable way~\cite{LBM+22}. This is true across different size models of the same generation (e.g., 13B versus 175B GPT-3). In our experience, transferabilty is increasingly difficult as models undergo different fine-tuning practices (e.g., instruction-tuning~\cite{WBZ+22}, alignment~\cite{OWJ+22} and function calling~\cite{SDD+23}). In two of our unreported experiments using $P_{G}$ with Meta's LLaMA2 base model and OpenAI's GPT-3 \texttt{davinci-002} base model, we found overall mean accuracy falls below 50\%, indicating that LM fine-tuning improves results, considerably.

\section{Conclusion and Future Work}
\label{section:conclusion}

In this paper, we report on experiments to check specifications for requirements satisfiability using large language models and in-context learning. We studied eight requirements from the legal domain using two popular models, GPT-3.5 and GPT-4. The experiments were conducted using synthetic data generated from popular mobile app descriptions that we acquired from jurisdictions regulated by this domain. In addition to different prompt templates with and without chain-of-thought (CoT) prompting, we studied LM performance when the specification both satisfies and dissatisfies the requirement and when the requirement is written in a normative and inverted tone. The results indicate that a generic prompt template outperforms a requirements theory-specific template, that CoT prompting improves mean accuracy well above non-CoT prompting with GPT-3.5, and that GPT-4 outperforms all other approaches studied.

The research highlights a number of challenges for future work. First, more work is needed to study the relationship between natural language and logical inference and to understand how these two approaches can complement one another. Because knowledge evolves, we need new tools to critique existing satisfaction arguments that no longer hold under changing requirements definitions. Finally, we see the ability to generate specifications (i.e., generative requirements engineering) as a rich opportunity for design space exploration. However, we need methods to engage humans in the analysis and comprehension of generative RE that can leverage and build on emerging work in reinforcement learning with human feedback (RL4HF)~\cite{CLB+17}.

\section*{Acknowledgment}

This research was funded by NSF Awards \#2007298, \#2217572, \#2217573, and NSF CAREER \#2238047.

\appendices

\section{Prompts to Generate Scenarios}

The scenario generation process consists of two stages: (a) summarize the app description; (b) identify actions from the app description; (c) identify data types for each action; (d) generate a scenario from the summary, actions and data types; and (d) generate specifications from the scenarios and property description. For each of the following prompts, we used the OpenAI API with the \textit{gpt-3.5-turbo-1106} model and parameters $temperature = 0.7$, $top\_p = 1.0$. We provide a single system message: ``You are a helpful assistant.''

Prior to summarization, all app descriptions are translated to English using the Google Translate API. In Listing~\ref{listing:summary_app}, we present the prompt template for summarizing the app description. The template accepts one parameter, which is the original app description obtained from each of the public app pages provided by the Apple App and Google Play stores.

\begin{minipage}{.95\linewidth}
\begin{lstlisting}[language=HTML,caption={Summarize App Description},label={listing:summary_app},basicstyle=\scriptsize\ttfamily]
Summarize the following app description in one sentence. Refer to the app as 'The mobile app' and do not refer to the app's name. Do not comment or elaborate.

App Description: {app_desc}
\end{lstlisting}
\end{minipage}

In Listing~\ref{listing:identify_actions}, we present the prompt to extract user and app actions from the original app description. For each action, we use the prompt in Listing~\ref{listing:identify_datatypes} to identify potential data types for the action. To illustrate the results of these two prompts, we present five randomly selected user actions and associate data types in square brackets, below.

\begin{enumerate}
    \item ``The user customizes units for speed, altitude, visibility, temperature, and atmospheric pressure.'', ['preferred units', 'location']

    \item ``The user creates a customized package of their favorite content to enjoy on all their devices.'', ['content preferences', 'devices used', 'viewing habits']

    \item ``The user can create their own user account to track Aki Awards, unlocked accessories, and Genizs' balance.'', ['user account information', 'Aki Awards progress', 'unlocked accessories', 'Genizs balance']

    \item ``The user learns a new language through Duolingo's educational app.'', ['language learning progress', 'lessons completed', 'time spent on exercises', 'correct/incorrect responses']

    \item ``The user saves all user credentials in the Keychain for convenience.'', ['user credentials']
\end{enumerate}

\begin{minipage}{.95\linewidth}
\begin{lstlisting}[language=HTML,caption={Identify User and App Actions},label={listing:identify_actions},basicstyle=\scriptsize\ttfamily]
Identify the main actions described in the following app description, and respond with a list in JSON format where each item in the list is a user or app action description. Each action should have a subject idenitfying who performs the action. Do not comment or elaborate. The JSON output should look like the following example.
    
Example: ["The user reads and responds to posts by their friends.", "The app provides support for multiple file formats.", "The user creates a video from their favorite dance moves.", "The app instructs the user how to speak in a foreign language."]

App Description: {app_desc}
\end{lstlisting}
\end{minipage}

\begin{minipage}{.95\linewidth}
\begin{lstlisting}[language=HTML,caption={Identify Data Types},label={listing:identify_datatypes},basicstyle=\scriptsize\ttfamily]
Identify the main actions described in the following app description, and respond with a list in JSON format where each item in the list is a user or app action description. Each action should have a subject idenitfying who performs the action. Do not comment or elaborate. The JSON output should look like the following example.
    
Example: ["The user reads and responds to posts by their friends.", "The app provides support for multiple file formats.", "The user creates a video from their favorite dance moves.", "The app instructs the user how to speak in a foreign language."]


App Description: {app_desc}
\end{lstlisting}
\end{minipage}

Prior to generating the scenario, we remove any actions that do not include at least one data action, and we remove any summaries that do not include at least one action. For example, in the case of 200 randomly selected top Apple App descriptions, we removed five descriptions that did not generate any actions with data types. We generate scenarios using the prompt in Listing~\ref{listing:rewrite_scenarios}, which requires the brief app summary, the data action and associated data types. For the Apple App scenarios, this step yielded 1156 scenarios, noting that each app description yielded multiple actions in the earlier steps.

\begin{minipage}{.95\linewidth}
\begin{lstlisting}[language=HTML,caption={Rewrite Actions and Data Scenario},label={listing:rewrite_scenarios},basicstyle=\scriptsize\ttfamily]
Rewrite the following sentences into a brief user scenario in third person. Refer to the user or app, but do not mention names. Do not mention the name of the app. Minimize overly expressive language.
    
Information: {summary} {action} The app uses {datatypes} to perform this function.

Scenario: 
\end{lstlisting}
\end{minipage}

Finally, Listing~\ref{listing:gen_spec} presents the prompt that accepts a generated scenario, property name and definition (see Section~\ref{subsection:knowledge}) and generates a list of actions that extend the scenario to either satisfy (\textit{state}=``true'') or not satisfy (\textit{state}=``false'') the given property.

\begin{minipage}{.95\linewidth}
\begin{lstlisting}[language=HTML,caption={Generate Specifications},label={listing:gen_spec},basicstyle=\scriptsize\ttfamily]
Carefully read the definition, and extend the specification to describe actions by the app and user that cause "{property}" to be {state}. Ensure each action in the extension supports that "{axiom}" is {state}. Do not refer directly to "{property}" in your response.
    
Definition of {property}: {definition}
    
Specification: {scenario}

Actions:
\end{lstlisting}
\end{minipage}

As an illustration, we present the following inputs (property name, state and scenario) to the above prompt and the resulting scenario extension generated using these inputs:

\textbf{Property Name}: Withdrawal

\textbf{Property State}: True

\textbf{Scenario}: A user who is a pilot needs to check the weather conditions for their upcoming flight. They open the app and easily access visualized METAR and TAF weather information for the airport they will be departing from. The user appreciates the easy-to-understand metrics and weather symbols provided by the app, which help them make informed decisions about their flight plan. They also find the crosswind calculations and flight rule categories helpful. After customizing the units to their preference, the user listens to the voice synthesis reading of the weather details and notes the sunrise and sunset times for their journey. The app utilizes the user's location and weather preferences to deliver accurate and personalized weather updates, making it a valuable tool for aviation weather forecasting.

\textbf{Scenario Extension}: 
\begin{enumerate}
    \item The app allows the user to easily customize their consent preferences for data processing related to weather updates by providing a dedicated settings section that clearly outlines the options for opting in or out of specific data processing activities.
    \item The user can access a clear and concise privacy policy within the app that details how their data is used for weather forecasting purposes and explicitly states their right to withdraw consent at any time.
    \item Upon opening the app, the user receives a notification reminding them of their right to withdraw consent for data processing, with a direct link to the settings where they can easily make changes to their consent preferences.
    \item The app automatically logs the timestamp and type of consent given by the user, ensuring that the withdrawal process is as seamless and transparent as possible.
    \item The user can easily navigate to a dedicated section within the app where they can review and manage their consent settings, including the option to withdraw consent with just a few taps or clicks.
    \item In case the user decides to withdraw consent, the app immediately stops processing their data for weather updates and notifies the user of the successful withdrawal.
    \item The app periodically reminds the user of their right to withdraw consent and provides helpful tips on how to easily adjust their settings if needed.
    \item The user receives an email confirmation after withdrawing consent, reassuring them that their decision has been successfully processed and their data will no longer be used for weather forecasting purposes.
    \item The app ensures that the withdrawal process is not hidden or buried within complex menus, making it easy for the user to find and exercise their right to withdraw consent at any given time.
    \item The user can easily re-enable consent for data processing at any time through a straightforward process within the app, ensuring that they have full control over their privacy preferences.
\end{enumerate}

\section{Complete Rubric and Property Definitions}

\textbf{Power imbalance} generally occurs when the data controller is a public authority or employer, although other cases may arise. For consent to be freely given in the presence of a power imbalance, the controller must demonstrate that there is no detriment when consent is refused or later withdrawn. Recital 43 clearly indicates that it is unlikely that public authorities can rely on consent for processing as whenever the controller is a public authority, there is often a clear imbalance of power in the relationship between the controller and the data subject. There may be situations when it is possible for the employer to demonstrate that consent actually is freely given. Given the imbalance of power between an employer and its staff members, employees can only give free consent in exceptional circumstances, when it will have no adverse consequences at all whether or not they give consent. In addition, Article 88 and Recital 155 describe the need to protect employee’s interests in order to avoid a power imbalance.

Requirement: There is a power imbalance between the data subject and the data controller

Req (Inverted): There is no power imbalance between the data subject and the data controller

\textbf{Conditionality}: If the purpose for processing a data type is bundled with other contract terms, or if the data subject is otherwise compelled to consent, then it is conditional and is not freely given. Conditionality only applies if the requested data is unnecessary to perform the contract. Contracts include end user agreements, terms of use, and terms and conditions. Article 7(4) GDPR indicates that, inter alia, the situation of “bundling” consent with acceptance of terms or conditions, or “tying” the provision of a contract or a service to a request for consent to process personal data that are not necessary for the performance of that contract or service, is considered highly undesirable.” Par 32. “Article 7(4) is only relevant where the requested data are not necessary for the performance of the contract, (including the provision of a service), and the performance of that contract is made conditional on the obtaining of these data on the basis of consent. Conversely, if processing is necessary to perform the contract (including to provide a service), then Article 7(4) does not apply.

Requirement: The data subject is compelled to consent or the purpose for data processing is bundled with other contract terms, such as user agreements, terms of use, or terms and conditions

Req (Inverted): The data subject is not compelled to consent and the purpose for data processing is not bundled with other contract terms, such as user agreements, terms of use, or terms and conditions

\textbf{Granularity}: Data subjects should be free to choose which purpose they accept, rather than having to consent to a bundle of processing purposes. Recital 43 clarifies that consent is presumed not to be freely given if the process/procedure for obtaining consent does not allow data subjects to give separate consent for personal data processing operations respectively (e.g. only for some processing operations and not for others) despite it being appropriate in the individual case. Recital 32 states, “Consent should cover all processing activities carried out for the same purpose or purposes. When the processing has multiple purposes, consent should be given for all of them.

Requirement: The data subject can choose which data processing purposes they accept

Req (Inverted): The data subject cannot choose which data processing purposes they accept

\textbf{Detriment}: The controller needs to demonstrate that it is possible to refuse or withdraw consent without detriment, including no deception, intimidation, coercion or significant negative consequences. Gray Area: permissible incentives, which means a controller can use an incentive that is only obtainable if the data subject consents. This incentive is not viewed as a detriment to refusing to consent. Refusal to consent or withdrawal should not lead to a diminished product or service. The controller needs to demonstrate that it is possible to refuse or withdraw consent without detriment ([see Recital 42]). For example, the controller needs to prove that withdrawing consent does not lead to any costs for the data subject and thus no clear disadvantage for those withdrawing consent.

Requirement: The data subject can withdraw consent and incur no detriment

Req (Inverted): The data subject may incur detriment if they withdraw consent

\textbf{Specificity}: The processing of data is limited to specific purposes and will not be processed for other purposes, the consent is granular, and the information presented to obtain consent describes the consent and not other unrelated matters. Article 6(1)(a) confirms that the consent of the data subject must be given in relation to “one or more specific” purposes and that a data subject has a choice in relation to each of them… In sum, to comply with the element of "specific" the controller must apply: i. Purpose specification as a safeguard against function creep, ii. Granularity in consent requests, and iii. Clear separation of information related to obtaining consent for data processing activities from information about other matters.

Requirement: Data processing is limited to specific purposes

Req (Inverted): Data processing is not limited to specific purposes

\textbf{Informed}: A design description must indicate that a data subject is informed prior to the collection of their data, and at minimum[9] identify (a) the data controller’s identity, (b) the purpose of each processing operation, (c) what type(s) of data will be collected and used, (d) the existence of the right to withdraw consent, (e) information about the use of the data for automated processing, and (f) about the risks due to transfers to countries without adequacy decisions or safeguards. Based on Article 5 of the GDPR, the requirement for transparency is one of the fundamental principles, closely related to the principles of fairness and lawfulness. Providing information to data subjects prior to obtaining their consent is essential in order to enable them to make informed decisions, understand what they are agreeing to, and for example exercise their right to withdraw their consent. For consent to be informed, it is necessary to inform the data subject of certain elements that are crucial to make a choice. Therefore, the EDPB is of the opinion that at least the following information is required for obtaining valid consent: i. the controller’s identity, ii. the purpose of each of the processing operations for which consent is sought, iii. what (type of) data will be collected and used, iv. the existence of the right to withdraw consent, v. information about the use of the data for automated decision-making in accordance with Article 22 (2)(c) where relevant, and on the possible risks of data transfers due to absence of an adequacy decision and of appropriate safeguards as described in Article 46.

Requirement: The data subject is properly informed prior to the collection of their data
Req (Inverted): The data subject is not property informed prior to the collection of their data

\textbf{Unambiguous}: Consent must be provided through a clear, affirmative action, which may be a written, oral or electronic means. Article 2(h) of Directive 95/46/EC described consent as an “indication of wishes by which the data subject signifies his agreement to personal data relating to him being processed”. Article 4(11) GDPR builds on this definition, by clarifying that valid consent requires an unambiguous indication by means of a statement or by a clear affirmative action, in line with previous guidance issued by the WP29. A “clear affirmative act” means that the data subject must have taken a deliberate action to consent to the particular processing. Recital 32 sets out additional guidance on this. Consent can be collected through a written or (a recorded) oral statement, including by electronic means.

Requirment: Consent is provided through a clear, affirmative action by the data subject

Req (Inverted): Consent is not provided through a clear, affirmative action by the data subject
 
\textbf{Withdrawal}: The data subject can withdraw consent as easily as they gave it, and at any given time. Article 7(3) of the GDPR prescribes that the controller must ensure that consent can be withdrawn by the data subject as easy as giving consent and at any given time. The GDPR does not say that giving and withdrawing consent must always be done through the same action.”, “However, when consent is obtained via electronic means through only one mouse-click, swipe, or keystroke, data subjects must, in practice, be able to withdraw that consent equally as easily.

Requirement: the data subject can withdraw consent as easily as they gave it and at any time

Req (Inverted): he data subject cannot withdraw consent as easily as they gave it

\vspace{12pt}
\end{document}